\documentclass[twocolumn]{aastex631}

\usepackage{color}
\usepackage{url}
\usepackage{hyperref}
\usepackage{xspace}
\usepackage{soul}
\usepackage{hyperref}
\usepackage{threeparttable}

\newcommand{\source}{\mbox{Swift~J1727.8$–$1613}\xspace}
\newcommand{\ixpe}{\textit{IXPE}}

\newcommand{\nustar}{\textit{NuSTAR}}
\newcommand{\nicer}{NICER}
\newcommand{\maxi}{MAXI}

\shorttitle{Swift~J1727.8$–$1613 in the soft spectral state}
\begin{document}

%\title{Substantial Decrease of the X-ray Polarization in the Soft Spectral State in Swift~J1727.8$–$1613}
%\title{Swift~J1727.8$–$1613 X-ray Polarization in the soft spectral state}
%\title{Dramatic Drop in the X-Ray Polarization of Swift~J1727.8$–$1613 in the Soft Spectral State: \\ Probing the Accretion Geometry}
\title{Dramatic Drop in the X-Ray Polarization of Swift~J1727.8$–$1613 in the Soft Spectral State}

%the author list

\author[0000-0003-2931-0742]{Ji\v{r}\'{i} Svoboda\thanks{E-mail: jiri.svoboda@asu.cas.cz}}
\affiliation{Astronomical Institute of the Czech Academy of Sciences, Bo\v{c}n\'{i} II 1401/1, 14100 Praha 4, Czech Republic}

\author[0000-0003-0079-1239]{Michal Dov\v{c}iak}
\affiliation{Astronomical Institute of the Czech Academy of Sciences, Bo\v{c}n\'{i} II 1401/1, 14100 Praha 4, Czech Republic}

\author[0000-0002-5872-6061]{James F. Steiner}
\affiliation{Center for Astrophysics $\vert$ Harvard \& Smithsonian, 60 Garden Street, Cambridge, MA 02138, USA}

\author[0000-0002-3638-0637]{Philip Kaaret}
\affiliation{NASA Marshall Space Flight Center, Huntsville, AL 35812, USA}

\author[0000-0001-5418-291X]{Jakub Podgorný}
\affiliation{Astronomical Institute of the Czech Academy of Sciences, Bo\v{c}n\'{i} II 1401/1, 14100 Praha 4, Czech Republic}

\author[0000-0002-0983-0049]{Juri Poutanen}
\affiliation{Department of Physics and Astronomy, FI-20014 University of Turku, Finland}

\author[0000-0002-5767-7253]{Alexandra Veledina}
\affiliation{Department of Physics and Astronomy, FI-20014 University of Turku, Finland}
\affiliation{Nordita, KTH Royal Institute of Technology and Stockholm
University, Hannes Alfv\'ens v\"ag 12, SE-10691 Stockholm, Sweden}

\author[0000-0003-3331-3794]{Fabio Muleri}
\affiliation{INAF Istituto di Astrofisica e Planetologia Spaziali, Via del Fosso del Cavaliere 100, 00133 Roma, Italy}	

\author[0000-0002-1768-618X]{Roberto Taverna}
\affiliation{Dipartimento di Fisica e Astronomia, Universit\`{a} degli Studi di Padova, Via Marzolo 8, 35131 Padova, Italy}

\author[0000-0002-1084-6507]{Henric Krawczynski}
\affiliation{Physics Department, McDonnell Center for the Space Sciences, and Center for Quantum Leaps, Washington University in St. Louis, St. Louis, MO 63130, USA}

\author[0009-0004-1197-5935]{Ma\"{i}mouna Brigitte}
\affiliation{Astronomical Institute of the Czech Academy of Sciences, Bo\v{c}n\'{i} II 1401/1, 14100 Praha 4, Czech Republic}
\affiliation{Astronomical Institute, Faculty of Mathematics and Physics, Charles University, V Holešovičkách 2, Prague 8, 180~00, Czech Republic}

\author[0000-0001-5975-1026]{Sudeb Ranjan Datta}
\affiliation{Astronomical Institute of the Czech Academy of Sciences, Bo\v{c}n\'{i} II 1401/1, 14100 Praha 4, Czech Republic}

\author[0000-0002-4622-4240]{Stefano Bianchi}
\affiliation{Dipartimento di Matematica e Fisica, Universit\`{a} degli Studi Roma Tre, Via della Vasca Navale 84, 00146 Roma, Italy}

\author[0000-0002-3348-4035]{Teo~Mu\~noz-Darias}
\affiliation{Instituto de Astrof\'isica de Canarias, 38205 La Laguna, Tenerife, Spain}
\affiliation{Departamento de Astrof\'\i{}sica, Universidad de La Laguna, E-38206 La Laguna, Tenerife, Spain}

\author[0000-0002-6548-5622]{Michela Negro} 
\affiliation{Department of Physics and Astronomy, Louisiana State University, Baton Rouge, LA 70803, USA}

\author[0000-0001-5256-0278]{Nicole Rodriguez Cavero}
\affiliation{Physics Department, McDonnell Center for the Space Sciences, and Center for Quantum Leaps, Washington University in St. Louis, St. Louis, MO 63130, USA}

\author[0000-0002-5870-0443]{Noel Castro Segura}
\affiliation{Department of Physics, University of Warwick, Gibbet Hill Road, Coventry CV4 7AL, UK}

\author[0009-0005-6609-5852]{Niek Bollemeijer}
\affiliation{Anton Pannekoek Institute for Astronomy, Amsterdam, Science Park 904, NL-1098 NH, The Netherlands}

\author[0000-0003-3828-2448]{Javier A. Garc\'{i}a}
\affiliation{X-ray Astrophysics Laboratory, NASA Goddard Space Flight Center, Greenbelt, MD 20771, USA}
\affiliation{California Institute of Technology, Pasadena, CA 91125, USA}

\author[0000-0002-5311-9078]{Adam Ingram}
\affiliation{School of Mathematics, Statistics, and Physics, Newcastle University, Newcastle upon Tyne NE1 7RU, UK}

\author[0000-0002-2152-0916]{Giorgio Matt}
\affiliation{Dipartimento di Matematica e Fisica, Universit\`{a} degli Studi Roma Tre, Via della Vasca Navale 84, 00146 Roma, Italy}

\author[0000-0002-9633-9193]{Edward Nathan}
\affiliation{California Institute of Technology, Pasadena, CA 91125, USA}

\author[0000-0002-5270-4240]{Martin C. Weisskopf}
\affiliation{NASA Marshall Space Flight Center, Huntsville, AL 35812, USA}

\author[0000-0002-3422-0074]{Diego Altamirano}
\affiliation{School of Physics and Astronomy, University of Southampton, University Rd, Southampton SO17 1BJ, UK}

\author[0000-0002-9785-7726]{Luca Baldini}
\affiliation{Istituto Nazionale di Fisica Nucleare, Sezione di Pisa, Largo B. Pontecorvo 3, 56127 Pisa, Italy}
\affiliation{Dipartimento di Fisica, Universit\`{a} di Pisa, Largo B. Pontecorvo 3, 56127 Pisa, Italy}

\author[0000-0002-6384-3027]{Fiamma Capitanio}
\affiliation{INAF Istituto di Astrofisica e Planetologia Spaziali, Via del Fosso del Cavaliere 100, 00133 Roma, Italy}

\author[0000-0002-1532-4142]{Elise Egron}
\affiliation{INAF Osservatorio Astronomico di Cagliari, Via della Scienza 5, 09047 Selargius (CA), Italy}

\author[0000-0002-2791-5011]{Razieh Emami}
\affiliation{Center for Astrophysics $\vert$ Harvard \& Smithsonian, 60 Garden Street, Cambridge, MA 02138, USA}

\author[0000-0002-9705-7948]{Kun Hu}
\affiliation{Physics Department, McDonnell Center for the Space Sciences, and Center for Quantum Leaps, Washington University in St. Louis, St. Louis, MO 63130, USA}

\author[0009-0001-4644-194X]{Lorenzo Marra}
\affiliation{Dipartimento di Fisica e Astronomia, Universit\`{a} degli Studi di Padova, Via Marzolo 8, 35131 Padova, Italy}
\affiliation{Dipartimento di Matematica e Fisica, Universit\`{a} degli Studi Roma Tre, Via della Vasca Navale 84, 00146 Roma, Italy}

\author[0000-0003-4216-7936]{Guglielmo Mastroserio}
\affiliation{Dipartimento di Fisica, Universit\`{a} degli Studi di Milano, Via Celoria 16, I-20133 Milano, Italy}

\author[0000-0001-6061-3480]{Pierre-Olivier Petrucci}
\affiliation{Universit\'{e} Grenoble Alpes, CNRS, IPAG, 38000 Grenoble, France}

\author[0000-0003-0411-4243]{Ajay Ratheesh}
\affiliation{INAF Istituto di Astrofisica e Planetologia Spaziali, Via del Fosso del Cavaliere 100, 00133 Roma, Italy}

\author[0000-0002-7781-4104]{Paolo Soffitta}
\affiliation{INAF Istituto di Astrofisica e Planetologia Spaziali, Via del Fosso del Cavaliere 100, 00133 Roma, Italy}

\author[0000-0002-6562-8654]{Francesco Tombesi}
\affiliation{Tor Vergata University of Rome, Via Della Ricerca Scientifica 1, 00133 Rome, Italy}
\affiliation{INFN - Roma Tor Vergata, Via Della Ricerca Scientifica 1, 00133 Rome, Italy}

\author[0000-0001-9108-573X]{Yi-Jung Yang}
\affiliation{Graduate Institute of Astronomy, National Central University, 300 Zhongda Road, Zhongli, Taoyuan 32001, Taiwan}
\affiliation{Department of Physics, National Cheng Kung University, University Road, Tainan, Taiwan}
\affiliation{Laboratory for Space Research, The University of Hong Kong, Cyberport 4, Hong Kong}

\author[0000-0002-2268-9318]{Yuexin Zhang}
\affiliation{Center for Astrophysics $\vert$ Harvard \& Smithsonian, 60 Garden Street, Cambridge, MA 02138, USA}
\affiliation{Kapteyn Astronomical Institute, University of Groningen, P.O.\ BOX 800, 9700 AV Groningen, The Netherlands}

%\date{} % Activate to display a given date or no date (if empty),
         % otherwise the current date is printed 

\begin{abstract}
Black-hole X-ray binaries exhibit different spectral and timing properties in different accretion states. The X-ray outburst of a recently discovered and extraordinarily bright source, Swift J1727.8$-$1613, has enabled the first investigation of how the X-ray polarization properties of a source evolve with spectral state. The 2--8 keV polarization degree was previously measured by the Imaging X-ray Polarimetry Explorer (IXPE) to be $\approx 4\%$ in the hard and hard intermediate states. Here we present new IXPE results taken in the soft state, with the X-ray flux dominated by the thermal accretion-disk emission.
%, about two orders of magnitude fainter than at the peak of the outburst. 
We find that the polarization degree has dropped dramatically to $\lesssim 1\%$.  This result indicates that the measured X-ray polarization is largely sensitive to the accretion state and the polarization fraction is significantly higher in the hard state when the X-ray emission is dominated by up-scattered radiation in the X-ray corona. The combined polarization measurements in the soft and hard states disfavor a very high or low inclination of the system.
%and are consistent with $i\approx35\degr$.
\end{abstract}

%% Keywords should appear after the \end{abstract} command. 
%% The AAS Journals now uses Unified Astronomy Thesaurus concepts:
%% https://astrothesaurus.org
%% You will be asked to selected these concepts during the submission process
%% but this old "keyword" functionality is maintained in case authors want
%% to include these concepts in their preprints.

%\footnote{\url{http://www.latex-project.org/}}

\section{Introduction}
\label{sec:intro}

Accreting stellar-mass black hole X-ray binaries (BHXRBs) represent ideal laboratories to study physics under extreme conditions of strong gravity. They are among the brightest X-ray sources in our Galaxy and have thus been very promising targets for the X-ray polarization mission {Imaging X-ray Polarimetry Explorer}  \citep[\ixpe,][]{Weisskopf2022}.
Most BHXRBs are transients, characterized by short (weeks to months) periods of activity, during which they have been observed to show various spectral states, each characterized by distinct broadband spectral and timing properties \citep{Zdziarski2004, Done2007, Belloni2010}.

In the ``hard'' state, the spectrum displays a power-law shape, arising from multiple Compton scatterings of low-energy photons within a hot medium, known as a hot accretion flow or a corona. 
Polarization in this state can be related to the scattering processes in the optically thin, flattened medium \citep{SunyaevTitarchuk1985,Poutanen1996} and is aligned with the minor axis of this medium.
The polarization degree (PD) grows with the system inclination and energy and can achieve $\sim$15--20\% at high energies if the medium is viewed nearly edge-on.
The first \ixpe\ data on the hard-state systems revealed the polarization signatures are largely consistent with these expectations \citep{Krawczynski2022,Veledina2023,Ingram2024}.

The ``soft'' state spectrum resembles blackbody radiation and is attributed to a multi-temperature blackbody component originating from an optically thick, geometrically thin accretion disk \citep{Shakura1973, Novikov1973}. 
Polarization in this case may arise from multiple scatterings in the optically thick medium -- the disk atmosphere \citep{Chandrasekhar1960,Sobolev1963,Rees1975}.
In contrast to the optically thin case, the polarization is expected to be aligned with the disk plane in the case of pure electron scattering (complete ionization of the atmosphere), albeit in certain specific cases of non-zero absorption, polarization may become aligned with the disk axis \citep{LoskutovSobolev1979}. In such case, the PD is a growing function of inclination, reaching 11.7\% for the edge-on pure electron scattering atmosphere, but is generally expected to be smaller for the same viewing angle, as compared to the optically thin, hard-state case. 
%In addition, these calculations are made without the general relativistic effects, which have further depolarizing effects \citep{Dovciak2008, Schnittman2009}.

The accretion disk is believed to extend down to the innermost stable circular orbit (ISCO), below which matter freely falls toward the black hole. The spectral shape is closely linked to the radius of the ISCO and, consequently, to the spin of the black hole. Additionally, the black hole spin affects the properties of X-ray polarization \citep{Connors1980,Dovciak2008, Schnittman2009,Loktev2023} and thus, X-ray polarimetry serves as an independent tool for determining the black hole spin of BHXRBs in the soft state \citep{Dovciak2008, Taverna2020, Mikusincova2023}, which has been already applied to the \ixpe\ observations \citep{Svoboda2024, Marra2023}.

Previous {\ixpe} observations of BHXRBs in the soft state include \mbox{4U 1630$-$47} \citep{Kushwaha2023,Ratheesh2024}, \mbox{LMC~X-1} \citep{Podgorny2023}, \mbox{LMC~X-3} \citep{Svoboda2024}, and \mbox{4U~1957+115} \citep{Marra2023}. \mbox{LMC~X-1} is a system with a low inclination angle ($\approx$\,30\degr) and, as anticipated, it exhibited a low polarization degree with an upper limit at around 1\%. In contrast, \mbox{LMC~X-3} and \mbox{4U 1957+115} are systems with higher inclination ($\approx$\,70\degr), and they displayed a higher PD of about 2\%--4\%. These sources showed an expected level of polarization consistent with the semi-infinite electron-scattering atmosphere of the standard geometrically-thin optically-thick accretion disk and thus allowed for the first time to estimate the black hole spin via X-ray polarimetry.
However, polarization measurements of \mbox{4U~1630$-$47} in the soft state revealed a surprisingly high level of polarization for the estimated inclination ($i < 75\degr$), exceeding 8\%, and a significant increase of the PD with energy, posing challenges to this established scenario. The high X-ray polarization of this source was subsequently detected in the steep power-law state \citep{RodriguezCavero2023}.

The newly discovered BHXRB transient, {\source} (R.A. 17:27:43.31, DEC $-$16:12:19.23) was initially detected on 2023 August 24  \citep{GCN.34540,GCN.34544}, showing a remarkably intense outburst, peaking at around 7~Crab in the 2--20~keV energy range. A comprehensive evaluation of the source characteristics strongly supports its classification as a BHXRB from its X-ray spectrum \citep{ATel16210, ATel16217, Peng2024}, the identification of Type-C quasi-periodic oscillations \citep{ATel16215, ATel16219, ATel16247, Zhao2024}, and the detection of flat-spectrum radio emission indicative of a compact jet \citep{ATel16211,ATel16228}. The presence of the black hole in this X-ray binary is also suggested from the optical spectroscopy %with the GTC-10.4\,m telescope 
\citep{Mata2024}. They reported the detection of inflows and outflows during the outburst and also derived the distance of the source to be $D = 2.7 \pm 0.3$ kpc, with the value and uncertainty estimated using various empirical scalings.

{\ixpe} first observed this source on 2023 September~7, recording a PD  of $\approx 4\%$ \citep{Veledina2023}. The X-ray polarization was found to be roughly in the North-South direction with a polarization angle (PA) $\approx 2\degr$ East of North, which aligns with the sub-mm polarization \citep{ATel16230}, as well as with the optical polarization \citep{ATel16245}. Thus, the 2--8 keV polarization is aligned with the jet similarly to the case of  Cyg~X-1 \citep{Krawczynski2022} and NGC~4151 \citep{Gianolli2023}. Subsequent {\ixpe} observations (2023 Sep 7, Sep 16, Sep 27, and Oct 4), performed until the end of the source visibility window, have covered the transition towards the soft state \citep{Ingram2024}. The 2--8 keV PD decreased from $\approx 4\%$ to $\approx 3\%$ as the source transitioned through the hard intermediate state, but the PA remained the same within the measurement uncertainties of $\approx 2\degr$.

%VLBI...?
% Peng 2024 - InSight, Nustar and Nicer observation - suggest moderate inclinate 40 deg and high spin 0.98
% Zhao 2024 - QPO

This paper reports on the first X-ray polarization measurements of {\source} in the soft state. The data reduction is described in Section~\ref{sec:data}, the obtained results are presented in Section~\ref{sec:results} and discussed in Section~\ref{sec:discussion}.
%possibly close to the spectral transition to the low/hard state and quiescence given its current flux that is two orders of magnitude weaker than during the outburst. 

\section{Observations and Data Reduction}
\label{sec:data}

%{\ixpe} conducted two additional observations of Swift J1727.8--1613 between 2024 February 11 ??:?? UTC and 2024 February 13 ??:?? UTC (Observation 6, total live time of ?? ks) and between 2024 February 20 ??:?? UTC and 2024 February ?? ??:?? UTC (Observation 7, total live time of ?? ks). In contrast to the initial five observations (Observations 1--5) documented in \cite{Veledina2023} and \cite{Ingram2024}, the new observations 6 and 7 captured the source in the soft state. Figure \ref{fig:maxi} shows the evolution of X-ray characteristics of Swift J1727.8--1613 within the hardness-intensity diagram, tracing the outburst from its onset. The data were obtained from {\it MAXI} monitoring \citep{Maxi_2009}, excluding contamination by the nearby GX 9+9 source.

\subsection{IXPE}

The new {\ixpe} observations of {\source} in the soft state were taken on 2024 February 11--12 (ObsID: 03005701; UTC time between 2024 February 11 09:59:05 and 2024 February 12 22:11:52) and February 20--23 (ObsID: 03006001; UTC time between 2024 February 20 02:11:04 and 2024 February 23 11:13:56).
A 67\,ks cleaned exposure from the first observation and 151\,ks from the second one were acquired. The Level-2 data of the three \ixpe\ detectors \citep{Soffitta2021} were downloaded from the HEASARC archive and then filtered for source regions with the \texttt{xpselect} tool from the \textsc{ixpeobssim} software package version 31.0.1 \citep{Baldini2022}. The source extraction regions were selected as circles with a radius of 100\arcsec. No background subtraction was needed since the source was bright enough \citep{DiMarco2023}.
%, and the background regions as annuli with an inner radius of 140\arcsec\ and an outer radius of 280\arcsec.
%\st{ Polarization cubes were generated using the {\textbf{weighted}} \texttt{pcube} algorithm }\citep{Baldini2022}. 
We used \texttt{xselect} from \textsc{heasoft} version 6.33 to generate the weighted Stokes $I$, $Q$, and $U$ spectra with the command \texttt{extract "SPECT" stokes=SIMPLE}. To use the most up-to-date {\ixpe} responses (\texttt{arf} and \texttt{mrf}), we used the \textsc{heasoft} tool \texttt{ixpecalcarf} with the event and attitude files of each observation (with weight=2, i.e. simple weighting). Finally, we used \texttt{grppha} to rebin the Stokes parameters to 30 bins in the 2--8\,keV energy band. Eventually, however, we limited the {\ixpe} energy range to 2--6\,keV because of the spectral discrepancies above 6\,keV with the other X-ray instruments.

\begin{figure}
\centering
\includegraphics[width=\linewidth]{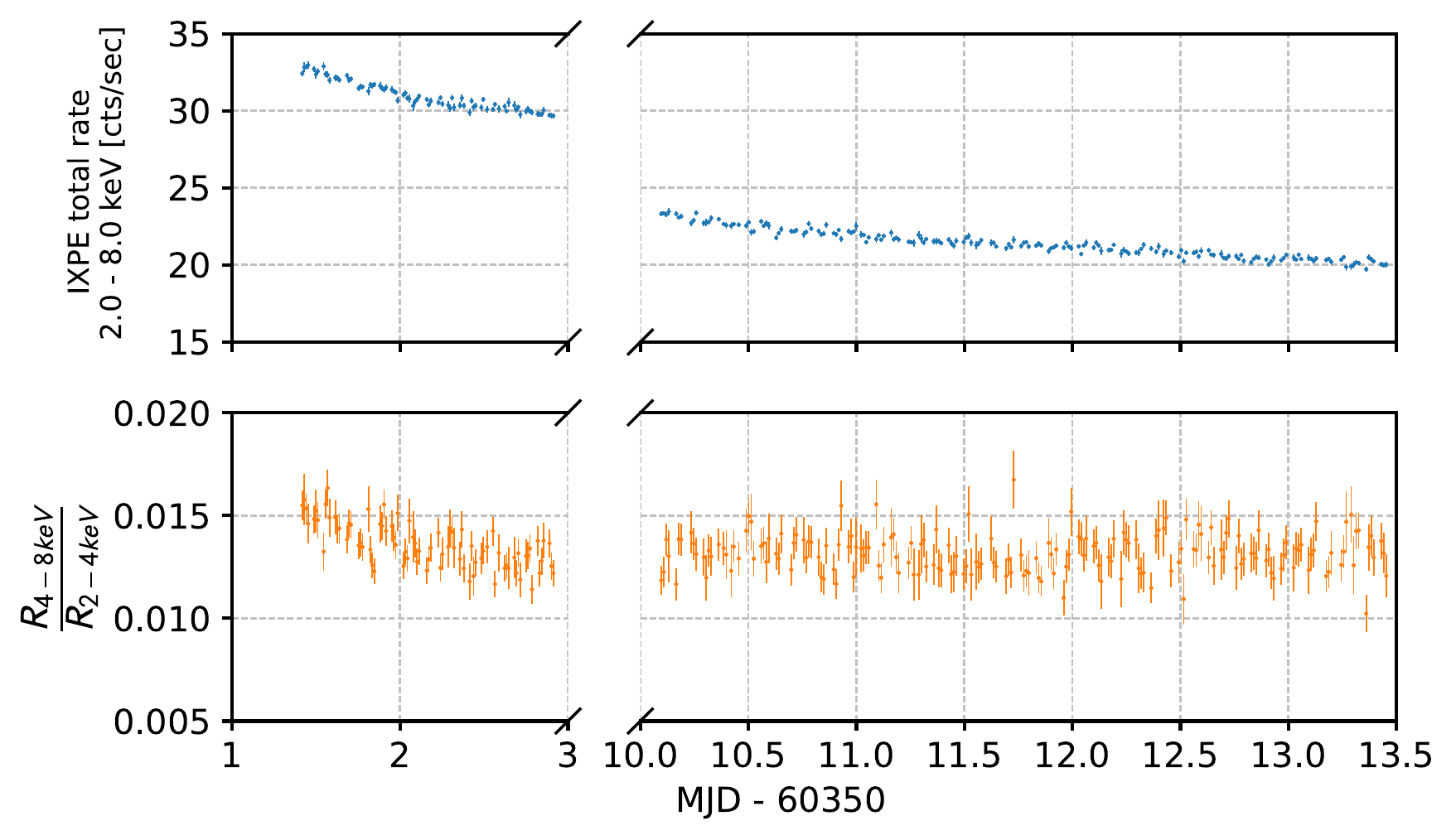}
\caption{Light curve between 2 and 8 keV of the two {\ixpe} observations, showing the decreasing count rate (upper panel) and roughly steady hardness (bottom panel) within measurement uncertainties during the observations. The time bin in both panels is 1~ks; hardness is defined as the ratio of the counting rate in the 4--8~keV and 2--4~keV energy ranges.} \label{fig:lc}
\end{figure}

The light curve of the two {\ixpe} observations is shown in Figure~\ref{fig:lc}, showing a continuous decrease of the count rate. Even while the X-ray flux dropped to about two thirds of its initial value at the beginning of the observation, the hardness stayed roughly constant within the measurement uncertainties, with a slight decreasing trend during the first observation.
The two {\ixpe} observations, being very similar, were merged into one for the spectro-polarimetric analysis so that we get stronger constraints on the polarimetric properties of the source. We performed merging with the \texttt{addspec} tool where the errors were propagated (\texttt{properr='yes' errmeth='Gauss'}) and new responses for merged Stokes spectra were computed. For this purpose, the original \texttt{rmf} responses had to be multiplied with \texttt{arf} or \texttt{mrf} into one \texttt{rsp} response file for $I$, $Q$, and $U$ Stokes spectra for each observation. We used the \texttt{marfrmf} tool for this task.

\begin{figure}
\centering
\includegraphics[width=\linewidth]{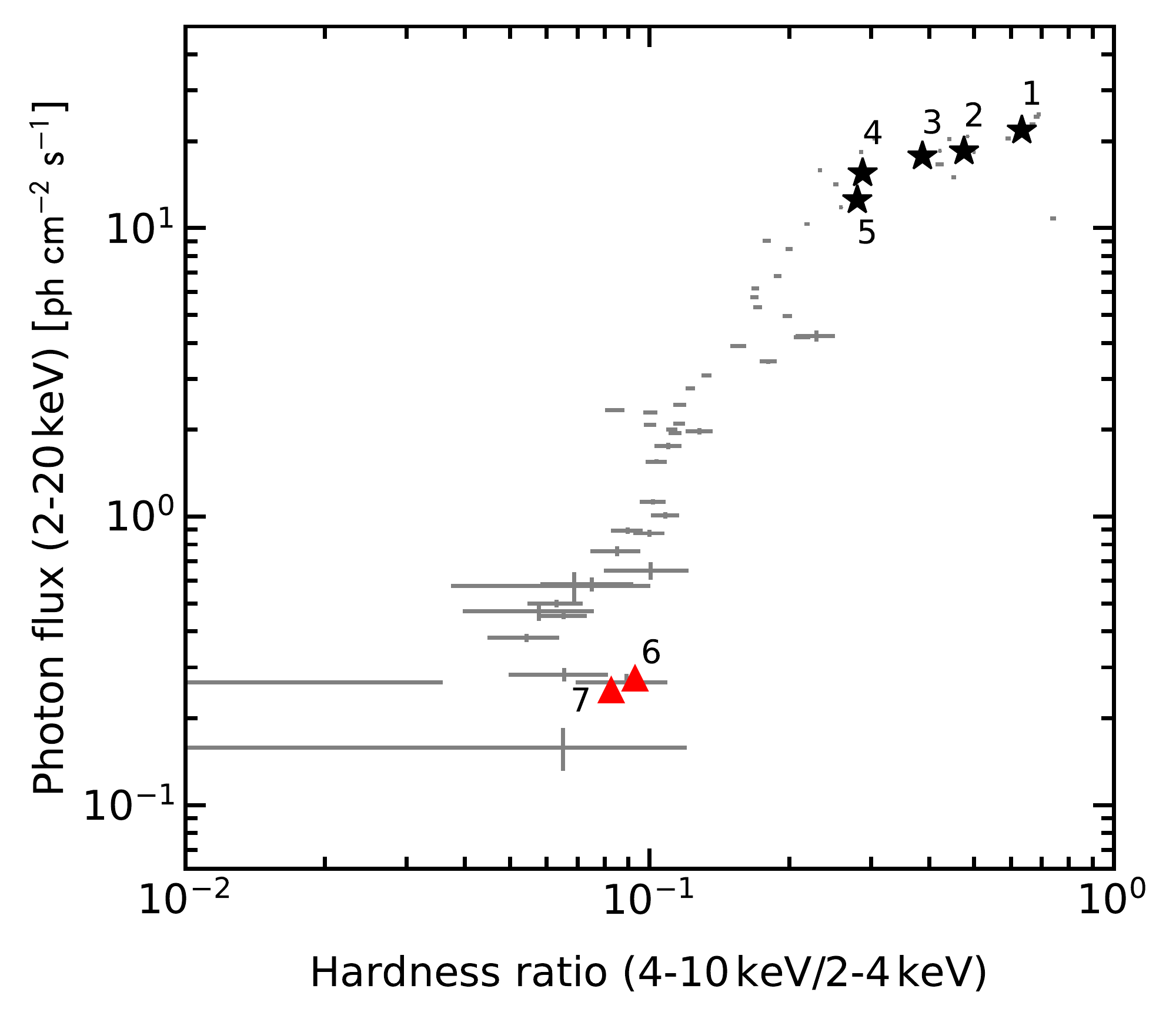}
\caption{Evolution of Swift J1727.8--1613 on the hardness-intensity diagram spanning from 2023 August 24 to 2024 February 23. The plot shows the ratio of the photon flux between the 4--10  and 2--4 keV bands against the photon flux in the 2--20 keV range, as determined from the MAXI ``on-demand'' archive with a default 3-day time bin. The black stars, labeled 1--5, denote the initial five observations by \ixpe\ that are described in \citet{Veledina2023} and \citet{Ingram2024}. The red triangles (labeled as 6 and 7) indicate the location in the diagram corresponding to the new \ixpe\ observations examined in this study, showing {\source} still in the soft state.} \label{fig:maxi}
\end{figure}

\subsection{MAXI}

{\source} has been regularly monitored by the MAXI X-ray instrument \citep{Maxi_2009}. Because of the MAXI wide field of view and the relative proximity (about 1\fdg2) of another bright X-ray source, a low-mass neutron star X-ray binary GX~9+9 \citep{Ursini2023}, we used the MAXI ``on-demand'' service for more precise delineation of the source region and background. The region around GX~9+9 was subtracted from both the source and background regions.

The MAXI data allow us to constrain the X-ray flux and spectral hardness and were used to trigger new X-ray observations of {\source} by confirming its presence in the soft state. The MAXI monitoring of {\source} up to the new \ixpe\  observation is summarized in Figure~\ref{fig:maxi}. The plot serves as an updated version of Figure~1 in \citet{Veledina2023} and \citet{Ingram2024} with slight variations due to the enhanced data selection, subtracting the region around GX 9+9.% The contamination

\subsection{NICER}

The {\nicer} data were obtained in 20 selected $\approx 500$~s exposures obtained under dark conditions on 2024 February 11--13 (ObsID: 6750010503-504, 7708010101-103). The data were reduced using {\sc nicerdas} software version 2023-08-22\_V011a. All observations were obtained with 50 active focal-plane modules (FPMs) out of {\nicer}'s 52 working FPM array.  Detectors 06 and 16 were manually excised out of precaution, based on elevated noise profiles noticed in earlier observations of \mbox{Swift J1727.8$-$1613} in daylight conditions (resulting in a $\sim 4\%$ loss of effective area).  The background was estimated for each exposure using the \texttt{scorpeon} model.  Several exposures with elevated background (flagged by when {\em both} overshoot rate $>$0.5\,s$^{-1}$ per FPM and cutoff-rigidity measure COR\_SAX$<$5) affecting high-energy sensitivity were excluded, as were any with short good-time intervals $<$60~s. Resultant spectral data were grouped to oversample the instrumental energy resolution by a factor $\sim 2.5$.
A systematic error of 0.5\% was added using the {\texttt{grppha}} tool.
%using the standard systematic error vector from the calibration database ({\textsc{syserrfile=caldb}}).

\subsection{NuSTAR}

{\nustar} observed {\source} on 2024 February 12 (ObsID: 80902348005, 2024-02-12T07:26:09 2024-02-12T19:56:09), i.e. simultaneously with the {\ixpe} observation. A 21.5\,ks net exposure time was obtained.
The {\nustar} data were reduced with the standard Data Analysis Software (\texttt{NuSTARDAS}). The {\nustar} calibration files from the CALDB database were used to calibrate the cleaned event files, produced by the \texttt{nupipeline} task.
Circular regions with a radius of 60\arcsec\  were centered on the source image for source extraction, and the background regions with radius 90\arcsec\ were selected from the corner of the same quadrant in the source-free region.
The source spectrum is rather soft and the background dominates over the source above 30\,keV. Therefore, we limit the {\nustar} data of \mbox{Swift J1727.8$-$1613} at high energies to be below 20\,keV in all spectral analysis to be sure the signal dominates over the background. Spectra from both the FPMA and FPMB detectors were used for the spectral analysis with a constant factor to account for cross-calibration uncertainties.

The data reduction is performed with the NASA's {\textsc{Heasoft}} software version 6.32.1. The \textsc{xspec} \citep{Arnaud1996}, version 12.13.1, is used for the spectral analysis.

\begin{table*}[t]
\centering
\footnotesize
\caption{Spectral fit parameters with the final spectral model.}
\begin{tabular}{cccccc}
\hline
\hline
%\rule{0cm}{0.3cm}
Component & Parameter  & Description & \multicolumn{3}{c}{Value}\\
& (units) & &
{\nicer} & {\nustar} & {\ixpe} \\ 
%\hline
\hline
%\texttt{constant} & & factor &  \multicolumn{3}{c}{$0.21 \pm 0.02$} \\
%\hline
\rule{0cm}{0.3cm}
\texttt{tbabs} & $N_{\rm H}$ ($10^{22}$\,cm$^{-2}$) & H column density &  \multicolumn{3}{c}{$0.24 \pm 0.01$} \\
\hline
\rule{0cm}{0.3cm}
{\texttt{kerrbb}} 
 & $M_{\rm bh}$ ($M_\odot$) & Black hole mass & \multicolumn{3}{c}{$10$ (frozen)}  \\ 
    & $a/M$ & Black hole spin & \multicolumn{3}{c}{$0.87 \pm 0.03$} \\
    & $\dot{M}$ [$10^{18}$\,g\,s$^{-1}$] & Accretion rate & $0.10 \pm 0.01 $ & $0.09 \pm 0.01 $ & $0.09 \pm 0.01 $\\
    & $i$ (deg) & Inclination & \multicolumn{3}{c}{$38 \pm 3$ } \\
%    & $D_\textrm{bh}$ (kpc) & Distance & \multicolumn{3}{c}{$2.7$ (frozen)} \\
 %   & hd & Color hardening  & \multicolumn{3}{c}{$1.7$} \\
 %   & $l_\textrm{flag}$ & Limb-darkening & \multicolumn{3}{c}{$1$ (frozen)} \\
 %   & $v_\textrm{flag}$ & Self-irradiation & \multicolumn{3}{c}{$1$ (frozen)} \\
 %   & norm & normalization & \multicolumn{3}{c}{$1$ (frozen)}  \\
\hline
\rule{0cm}{0.3cm}
{\texttt{thcomp}} & $\Gamma$ & Photon index &  \multicolumn{3}{c}{$4.9 \pm 0.3$} \\
%\multicolumn{3}{c}{$2.5^{\rm pegged}_{-0.2}$} \\ 
& CovFrac & covering fraction &  $0.02 \pm 0.01$ &  \multicolumn{2}{c}{$0.06 \pm 0.02$} \\
\hline
\rule{0cm}{0.3cm}
{\texttt{kynxillver}} & $L$ [$10^{-7}\,L_\mathrm{Edd}$] & Luminosity &  \multicolumn{3}{c}{$5 \pm 1$} \\
%\multicolumn{3}{c}{$2.5^{\rm pegged}_{-0.2}$} \\ 
%& norm & normalization &  \multicolumn{3}{c}{$0.0027^{+..}_{-..}$}\\
\hline
\rule{0cm}{0.3cm}
$\chi^2$/ dof & & & \multicolumn{3}{c}{340/329$\approx 1.03$} \\
%\hline
\hline
\end{tabular}
\label{table:spectral}
\begin{tablenotes}
\item {\bf{Notes:}} Quoted errors correspond to 90\% confidence levels. Further details on the model assumptions and modeling instrumental features are provided in Appendix~\ref{appendix-spectral}.
\end{tablenotes}
\end{table*}

\begin{figure} 
\centering
\includegraphics[width=\linewidth]{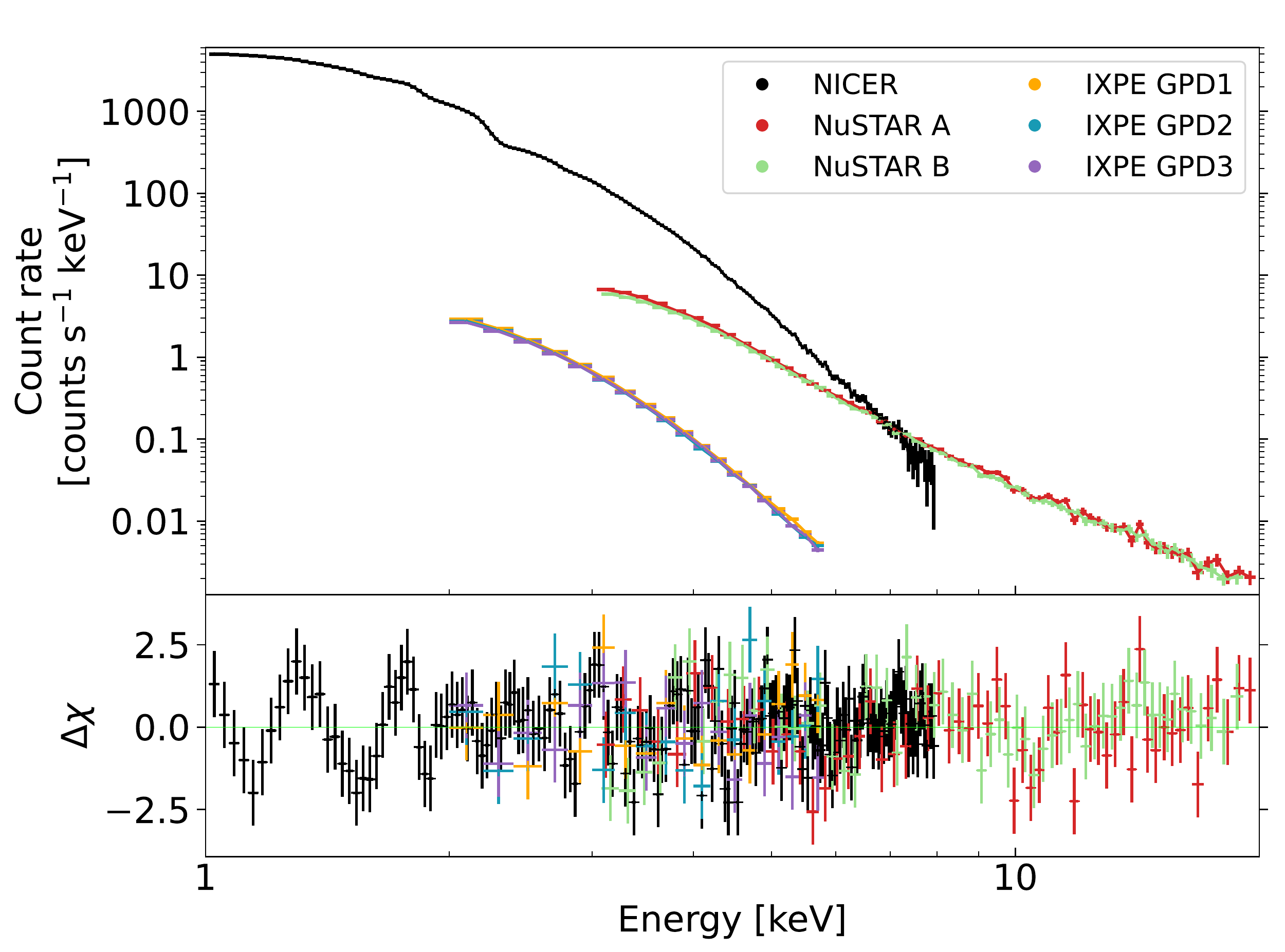}
\caption{Spectral fit of NICER (black), NuSTAR (FPMA in red, FPMB in green), and IXPE (orange, blue, and violet) data. The upper panel shows the data counts,
%the middle -- the deconvolved $EF_E$ spectrum (with \texttt{plot eemo} in \textsc{xspec}), 
and the bottom one data residuals from the best-fit model.}
 \label{fig:spectrum}
\end{figure}

\section{Results}
\label{sec:results}

The {\maxi} hardness ratio, defined as the ratio of the photon flux between the 4--10  and 2--4 keV, $H \approx 0.1$ depicts {\source} in the soft state (see Figure~\ref{fig:maxi}). This is confirmed by both the low hardness measured by {\ixpe}, and also from low variability - using {\nicer}, we constrained the 2--10 keV fractional rms in the 0.1--10 Hz to be $<2\%$, which is typical for the soft state \citep{2011MNRAS.410..679M}.

For the spectral analysis, we employed the {\nicer}, {\nustar}, and {\ixpe} observations. The {\texttt{tbabs}} model \citep{Wilms2000} was used to account for the line-of-sight absorption, with the column density as a free parameter. We included the relativistic model of the Novikov-Thorne accretion disk, \texttt{kerrbb} \citep{Li2005}, Comptonized by the thermal electrons modeled with \texttt{thcomp} \citep{Zdziarski2020}, and the relativistic reflection modeled with \texttt{kynxillver} as part of the \textsc{ky} code \citep{Dovciak2004}. We tied all the physical parameters, such as spin and inclination, between the thermal and reflection models.
Since there is a well-known degeneracy between the spin, inclination, mass, and distance in the X-ray continuum fitting method \citep{Remillard2006}, we fixed the distance to $D = 2.7$\,kpc \citep{Mata2024} and the black hole mass to 10 Solar masses.

The parameters were also tied between the {\nicer}, {\nustar}, and {\ixpe} observations except for the accretion rate and covering fraction of the Comptonization component. 
These were allowed to be free because the observations are not strictly simultaneous and, thus, these parameters can vary between the observations.
However, since the {\ixpe} spectra were not sensitive enough to the Comptonization model component, we fixed it to be the same as for {\nustar}. 
Similarly, the reflection model parameters were linked between all observations.
Details of spectral modeling are provided in Section~\ref{appendix-spectral}, and
the best-fit values of the physical parameters are reported in Table~\ref{table:spectral}. 

With our final spectral model, we obtained a perfectly acceptable fit with the reduced chi-squared value $\chi^2$/dof=340/329$\,\approx 1.03$. 
The data and the data residuals from the best-fit model are shown in Figure~\ref{fig:spectrum}. 
We obtained an absorption column density $N_{\rm H} = (0.24 \pm 0.01) \times 10^{22}$\,cm$^{-2}$, in agreement with the measured column density in this direction $N_{\rm  HI4PI} = 0.2 \times 10^{22}$\,cm$^{-2}$ from the HI4PI survey \citep{HI4PI}. 
The dominating component is the thermal emission with the accretion rate constrained as $\dot{M} = 10^{17}$\,g\,s$^{-1}$, with the Comptonization contributing only by a few percent. A discrepancy in the values for scattering fraction between {\nicer} and {\nustar} can be attributed to a non-simultaneity of both data sets in conjunction with small calibration differences.

%We measured the black hole mass to be $M_{\rm bh} \approx 17 M_\odot$, consistent with classifying the compact object as a black hole. 
The constrained black hole spin is $a \approx 0.9$ and the inclination $i \approx$\,40\degr. We note, however, that the spin and inclination angle are mutually degenerate, %\citep[see, e.g.,][]{Remillard2006}, 
and the systematic uncertainty in the determination of these values is much larger than the statistical errors reported in Table~\ref{table:spectral}. 
They are also dependent on the assumed mass and distance parameters. More precise constraints on the physical parameters will be possible after a more accurate determination of the mass, e.g. a dynamical mass from optical spectroscopy in quiescence. For the remainder of our analysis, the precise values of the degenerate quantities is of secondary consideration.  Of central importance, we have shown a spectral model which describes the broadband X-ray spectrum very well, and which unambiguously shows a dominant thermal multicolor disk component.

\begin{figure} 
\centering
\includegraphics[width=\linewidth]{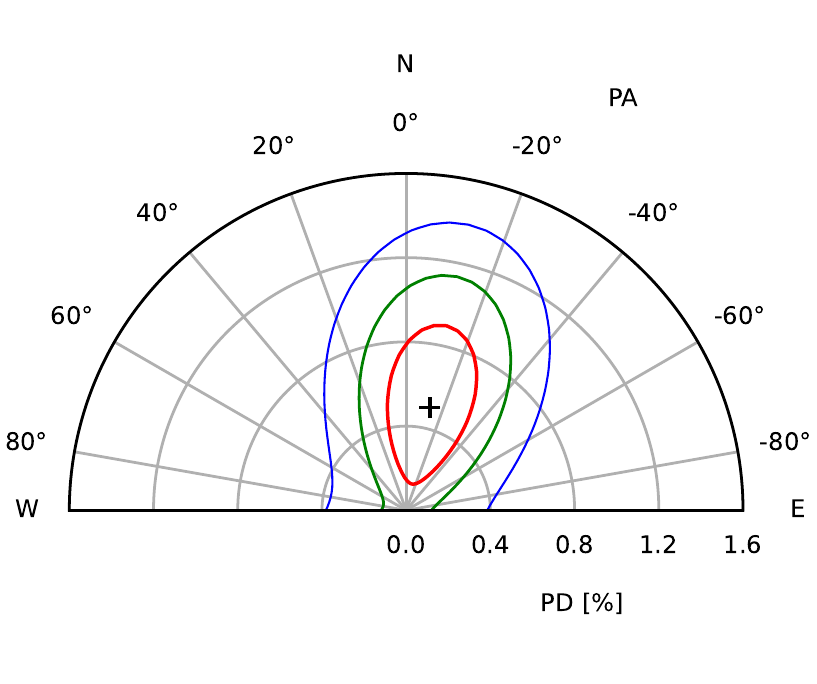}
\caption{Contour plot of the PD vs PA derived from the final spectral fit with applied {\texttt{polconst}} on the best-fit model to fit $Q$ and $U$ spectra. The contours show 1$\sigma$ (red, thick), 2$\sigma$ (green), and 3$\sigma$ (blue, thin) confidence levels. The best-fit values are marked with a black cross.} \label{fig:Polar}
\end{figure}

As a next step, we conducted the spectro-polarimetric analysis for the {\ixpe} $I$, $Q$, and $U$ spectra in the 2--6\,keV band using the {\texttt{polconst}} model applied to the best-fit spectral model with the parameters fixed to the values from the joint {\nicer}, {\nustar} and {\ixpe} fit. The same gain factors as for $I$ spectra were applied to $Q$ and $U$ spectra. The only free parameters were the PD and PA.
%, noted as $A$ and $\psi$ in {\texttt{polconst}}. 
We obtained the PD $=0.5^{+0.5}_{-0.4}\%$ and PA $= -13\degr \pm 28\degr$ with 90\% confidence errors. With 99\% confidence level, only an upper limit is constrained, PD\,$< 1.2\%$, and thus PA remains unconstrained. The fit is perfectly acceptable, with $\chi^2$/dof $= 168/170 \approx 1.0$. The contours for different confidence levels in the PD--PA plane were computed with 50 steps for each parameter and are shown in Figure~\ref{fig:Polar}. %The MDP level at 1.2\% is drawn in the plot by a black dashed line.
%The PA is unconstrained but not inconsistent with the original measurement in the hard state with the PA of $\approx 0\degr$.

\begin{figure}  
\centering
\includegraphics[width=\linewidth]{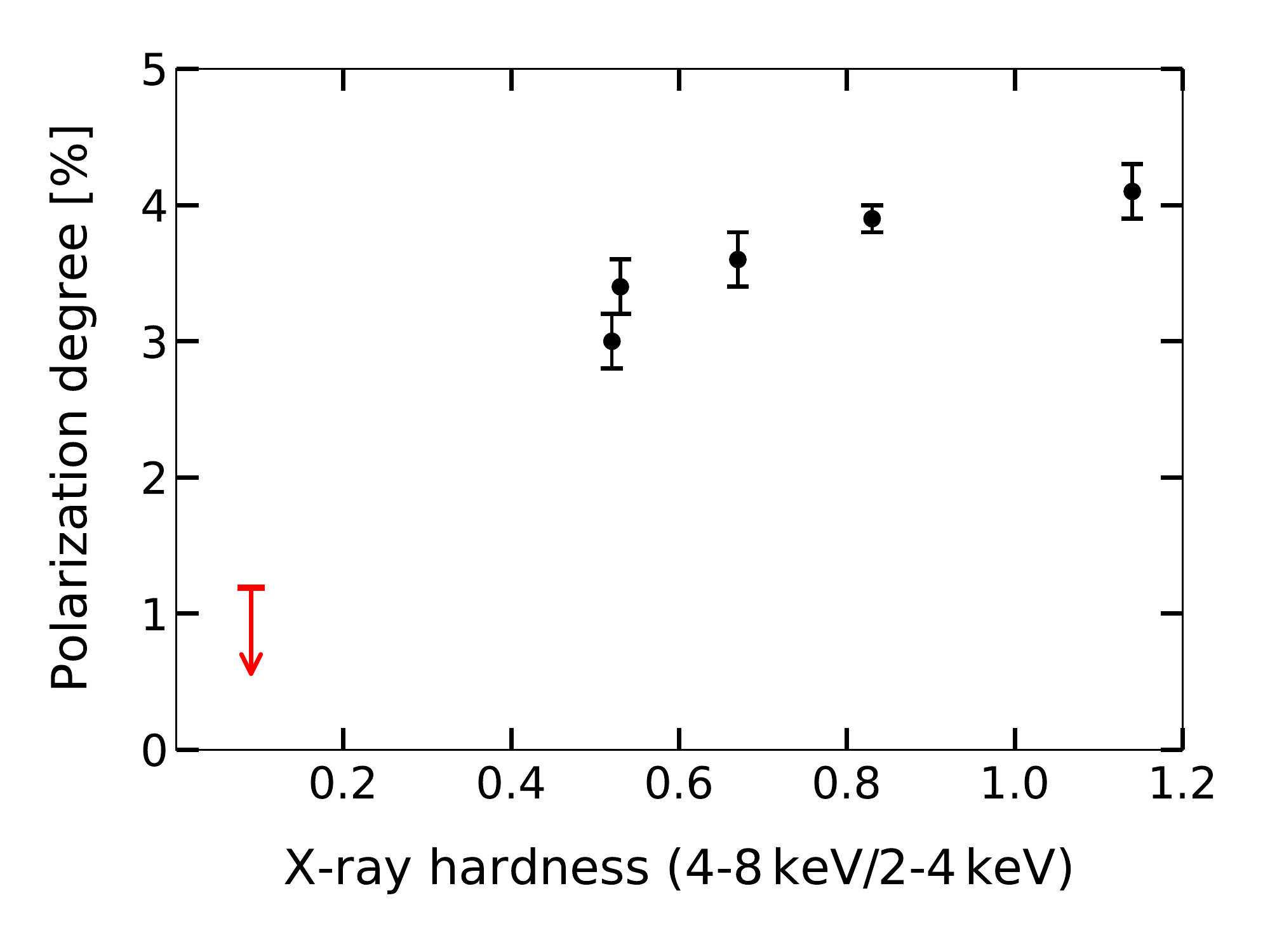}
\caption{Measured 2--8\,keV average PD in all {\ixpe} observations vs the spectral hardness defined as the flux ratio between the 4--8 and 2--4\,keV energy bands measured by {\ixpe}. The new observations 6 and 7 combined (marked in red) are plotted as an upper limit calculated at the 99\% confidence level.} \label{fig:pd_comparison}
\end{figure}

\section{Discussion}
\label{sec:discussion}

In this Letter, we report the first results of the X-ray spectro-polarimetric analysis of {\source} in the soft state about half a year after the beginning of the outburst. The MAXI hardness-intensity evolution of {\source} exhibits a monotonically declining intensity since the peak in the hard state; in February 2024 the intensity was about two orders of magnitude lower than the peak (see Figure~\ref{fig:maxi}).
%Between October 2023 and February 2024, the source was not visible for {\ixpe} due to the Sun constraints.
Our new observations showed that the X-ray polarization has substantially changed since the last observation in October 2023 in relation to the spectral state changes.
Figure~\ref{fig:pd_comparison} shows the PD measurements in the different observations as a function of the spectral hardness measured by {\ixpe}, defined as the ratio between the flux in the 4--8 and 2--4\,keV energy bands. The first observations from the beginning of the hard-to-soft state transition provided high values of PD and spectral hardness. As the source evolved and got softer, the PD was decreasing \citep{Ingram2024}, which is fully confirmed by the new measurement, with the spectral hardness $H \approx 0.1$ and PD$<1.2\%$.

%In the considered observations, however, the soft X-ray emission is still dominated by the thermal radiation of the accretion disk with $\gtrsim 90\%$ contributing to the X-ray flux in 2--8\,keV. 

The accretion geometry associated with the various spectral states and its possible evolution between them remain the subject of intense discussions \citep[see e.g.][]{Done2007}. 
Optically thin plasma is thought to be the main component responsible for the X-ray production in the hard state, while the soft state is thought to be related to the optically thick, geometrically thin disk emission. 
The states are likewise expected to have different polarization properties.

\begin{figure} 
\centering
\includegraphics[width=0.99\linewidth]{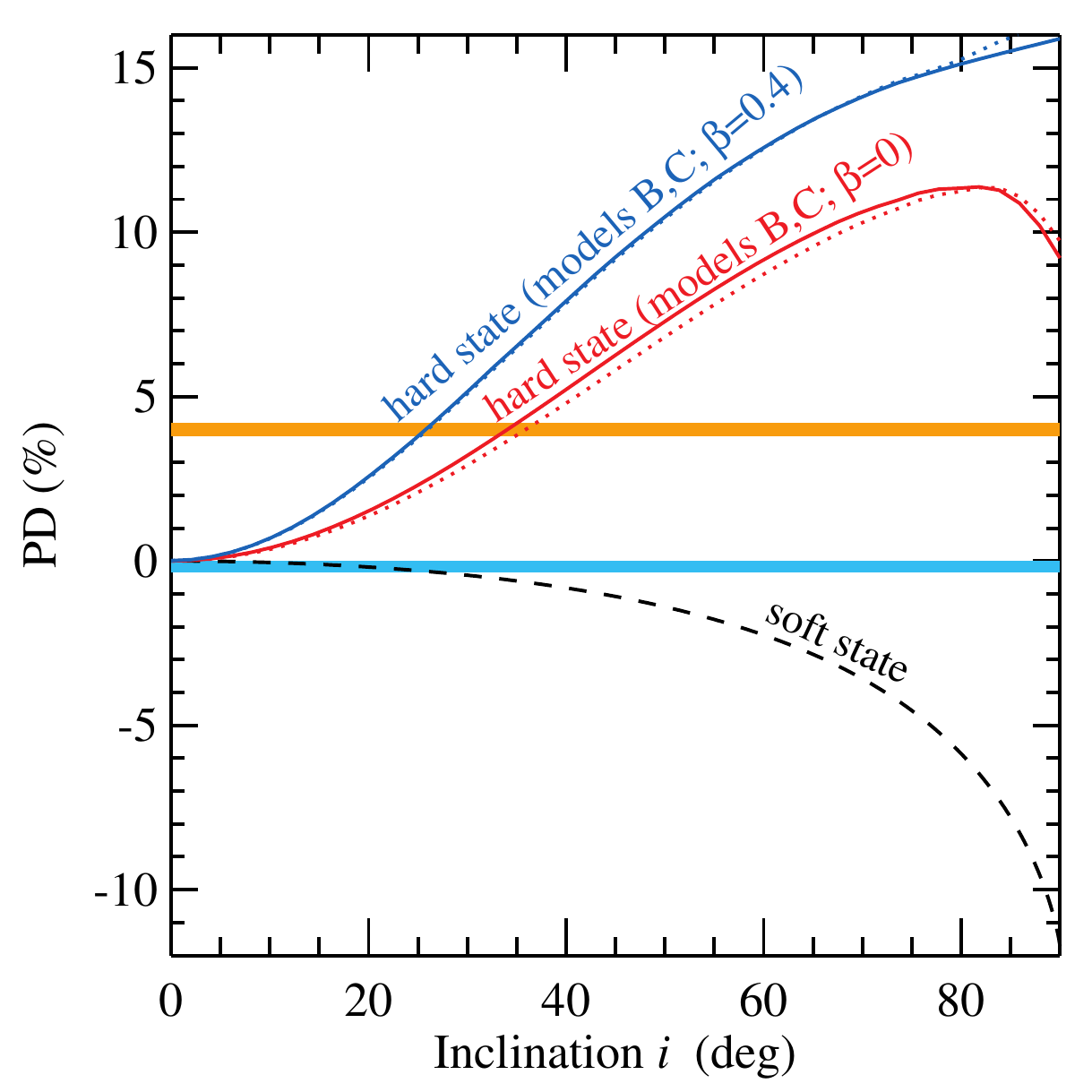}
\caption{Dependence of the PD on inclination. 
The red lines correspond to the models of Comptonization in the flat static hot flow with the seed photons either being internal synchrotron (solid line) or from the outer truncated disk (dotted line), corresponding to the models B and C in \citet{Poutanen2023}, respectively. 
The blue lines are for the same models but for the hot flow outflowing with velocity $v=0.4c$.  
The horizontal orange strip marks the observed PD of 4\% during the hard state. 
The inclinations in the range $\sim25\degr$--$35\degr$\ are consistent with this PD. 
The black dashed line is the classical result for the electron-scattering dominated semi-infinite atmosphere \citep{Chandrasekhar1960, Sobolev1963} that could correspond to the soft state. 
%The negative PD means that the PA is parallel to the disk.  
The horizontal cyan strip marks the allowed range of the PD during the soft state for the PA perpendicular to that of the hard state (i.e. corresponding to the negative PD). } \label{fig:PD_vs_incl}
\end{figure}

% Hard state polarization
% Complicatios due to various combinations of kT_eff and tau (for the same power-law slope), as well as exact shape/opening angle of the corona.
The hard-state polarization can be produced by multiple Compton scatterings in an optically thin, flat plasma cloud located either on the top of the disk or within its truncation radius \citep{SunyaevTitarchuk1985, Esin1997, Poutanen1997}.
The PD in this case is an increasing function of energy in the range between the first scattering and, roughly, the cut-off energy of the Comptonization continuum \citep[e.g.,][]{Poutanen1996}.
The PD generally increases with the inclination and also depends on the electron temperature and optical depth of the medium.
We first assume the model of a static flat hot flow Comptonizing either internal synchrotron photons or those coming from the truncated disk, and show the dependence of the resulting PD on inclination in the middle of the \ixpe\ range (at 4~keV)  in Figure~\ref{fig:PD_vs_incl} with red lines. 
These cases correspond to models B and C in \citet{Poutanen2023}.
The parameters, electron temperature $kT_{\rm e}=100$~keV, seed blackbody photon temperature $kT_{\rm bb}=0.3$~keV (for model C), and the photon index of the Comptonization component $\Gamma=1.8$, were chosen to match the hard-state data from \citet{Veledina2023}. 

The hot flow may also be outflowing, and therefore, we considered the case with the flow velocity of $v=0.4\,c$, which well modeled the Cyg X-1 data \citep{Krawczynski2022,Poutanen2023}. 
The models B and C with the outflow are shown with blue lines in Figure~\ref{fig:PD_vs_incl}. 
%The hard-state PD of \source favors the inclination higher than $\sim$\,25\degr--35\degr, depending on the outflow velocity. 
%A higher inclination is possible if the hot flow is not a perfect slab, which reduces the predicted PD.
The hard-state PD of Swift J1727.8–1613 favors the inclination around $\sim25\degr$--$35\degr$, depending on the outflow velocity. However, a higher inclination is possible if the hot flow is not a perfect slab, which reduces the predicted PD.
The slab-corona model (model A in \citealt{Poutanen2023}) with seed photons from the underlying cool disk predicts a strong energy dependence close to the energies of the seed photons, i.e. in the \ixpe\ range, which was, however, not observed in the hard-state \ixpe\ observations. 

\begin{figure*} 
\centering
\includegraphics[width=0.8\linewidth]{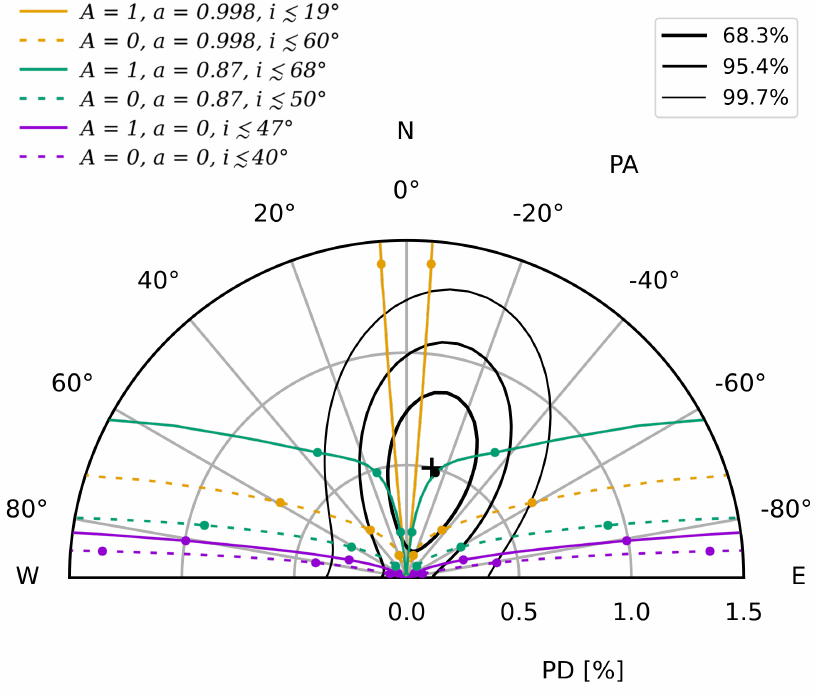}
\caption{Modeling PD and PA with relativistic accretion-disk model {\texttt{kynbbrr}} for different BH spin values, $a=0$ (violet), $a=0.87$ (green), and $a=0.998$ (orange), considering albedo $A=0$ (dashed lines) and $A=1$ (solid lines). Values for inclinations of $20\degr$, $40\degr$ and $60\degr$ are denoted by bullets. Corresponding upper limits for inclination are taken from the intersections with the 3\,$\sigma$ contour line. The black bullet shows the case of $a=0.87$ and inclination $i=40\degr$, corresponding to the best-fit spectral results, and it is very close to the best-fit values of PD and PA from the {\ixpe} data (denoted by a black cross).} \label{fig:Polar-model}
\end{figure*}

% Soft state polarization
% Some complications due to GR effect that affect mostly the soft-state polarimetric data (narrow peaks of bb emission vs power-law-like spectra from plasma at different radii)
The soft-state polarization can be attributed to Thomson scattering in the atmosphere of the optically thick accretion disk.
The observed upper limits on PD in \source can be compared to the classical \citet{Chandrasekhar1960} and \citet{Sobolev1963} results for the electron-scattering dominated semi-infinite atmosphere (Figure~\ref{fig:PD_vs_incl}, black dashed line).
In this case, the PD also increases with increasing inclination; however, the levels of polarization are expected to be 2--10 times lower, as compared to the optically thin case. The PA is rotated by 90\degr, i.e. the predominant direction of oscillations of the electric vector is aligned with the disk plane (and therefore, PD is shown as negative in Figure~\ref{fig:PD_vs_incl}). 
Assuming the PA perpendicular to that of the hard state, the upper limit on the PD is $\sim0.4\%$ (see Figure~\ref{fig:Polar}), which implies that in this model the inclination is limited by $i\lesssim30\degr$ (see the crossing of the cyan strip with the soft-state model in Figure~\ref{fig:PD_vs_incl}).   

The emission close to the BH experiences additional depolarization resulting from relativistic effects \citep{ConnorsStark1977,Connors1980}. However, at a high spin, the effects of returning radiation (i.e. thermal photons returning to the disk due to general-relativistic ray bending) become important \citep{Schnittman2009}.
%the self-irradiation effects become important \citep{Schnittman2009}.
Since, to be collected at infinity, returning photons must be first reflected on the disk surface, their polarization is expected to be aligned with the disk axis. As a result, if the fraction of returning photons reflected towards the observer is high (i.e. at a high albedo), the total polarization may also become aligned with the disk axis, even if the emission comes from the disk only.
%The polarization of returning radiation is aligned with the disk axis, and if its fraction is high in the total spectrum (i.e. at high albedo), the total polarization may also become aligned with the disk axis.
For relatively high spins of the black hole ($a\approx0.9$), this effect on X-ray polarization may become important even without a relevant contribution of the reflected photons being clearly visible in the 1--20\,keV spectrum \citep[see, e.g., Figure 5 in][]{Schnittman2009}.
This has been further studied in the standard Novikov-Thorne accretion disk scenario with additional self-irradiation for different combinations of inclination, black hole spin, and albedo parameters in the \texttt{kynbbrr} model \citep{Dovciak2008, Taverna2020, Mikusincova2023}.
%We note that all kinds of depolarization effects are more pronounced in the soft state, where the emission of each disk annulus has a peaked blackbody-like shape, further enhancing the contrast between the hard- and soft-state polarization. 

Figure~\ref{fig:Polar-model} shows how the polarization properties (PD and PA) vary with the observer's inclination according to the \texttt{kynbbrr} model, for different values of the black hole spin and the disk surface albedo.
%The PD and PA obtained with the {\texttt{kynbbrr}} model is shown in Figure~\ref{fig:Polar-model}. Different cases of the black hole spin and albedo (accounting for self-irradiation) with varying inclinations are considered there. 
The models are plotted over the contours of PD and PA obtained from the spectro-polarimetric analysis with the \texttt{polconst} model on the best-fit spectral model (Figure~\ref{fig:Polar}). The vertical north-direction axis is assumed to be the accretion disk axis, consistently with the hard-state polarization measurements. Since the orientation of the black hole spin and the direction of the accretion disk rotation cannot be determined a priori from the \texttt{kynbbrr} simulations, the models in Figure~\ref{fig:Polar-model} are plotted on both sides of the vertical axis.

The points of the intersections between the model curves and 3\,$\sigma$ contours represent limits for the inclination angle. The low-spin case ($a=0$) is close to Chandrasekhar's approximation and predicts the perpendicular PA direction. The allowed inclination is up to $40\degr$ for albedo $A=0$, and $47\degr$ for albedo $A = 1$. Oppositely, for the high spin ($a=0.998$) and high albedo ($A = 1$), the PA is oriented in the same direction as in the hard state, and the allowed range for inclination is even tighter ($i < 19\degr$). The albedo plays a crucial role in the high-spin scenario. For high spin and zero albedo, the PA is closer to the perpendicular direction and the inclination can be up to $60\degr$. The case of the BH spin $a=0.87$ corresponds to the best-fit value of the spectral fit. The allowed inclination for the albedo $A= 1$ is up to $68\degr$. Predicted PD and PA values assuming the inclination angle $i = 40\degr$ from the spectral fit (indicated by a black circle) are very close to the best-fit values of PD and PA from the polarimetric measurements.

The perfect agreement between the independent spectral and polarimetric analysis indicates consistency between these methods, though it can still be just coincidental given the relatively large uncertainties in our polarization measurements and systematic uncertainties in the X-ray continuum fitting method. 
Nevertheless, it is clear from this analysis that accounting for the general-relativistic effects allows the inclination over $30\degr$ to be still consistent with the soft-state PD. A more complicated picture can be obtained if we consider a possible contribution of the Comptonization or reflection components. These are relatively weak in the spectrum but can significantly more contribute to the polarization. The data sensitivity, however, does not allow us to perform such a detailed analysis, and we limited our analysis by assuming that the polarization is from the dominating thermal component.

Although the main focus of this Letter is on the X-ray polarization results, our spectral fit provides measurements of the physical parameters of the black hole in {\source}. The black hole spin derived from the joint X-ray continuum and reflection modeling provided $a \approx 0.9$, slightly lower but within uncertainties consistent with the value reported by \citet{Peng2024} from reflection modeling in the hard state. We note, however, that the accuracy of the spin measurements is, in general, affected by systematic uncertainty. The most limiting factor in our analysis is that the spin value is degenerate with other model parameters, and especially in the X-ray continuum fitting method, strongly correlates with values of the black hole mass, disk inclination, distance, the spectral hardening factor and also the disk structure \citep{Zdziarski2024}.
% A more comprehensive spectral analysis is beyond the scope of this paper and will be subject of a follow-up work.

The distance $D = 2.7$\,kpc constrained from the optical spectroscopy is larger than expected previously, simply estimated by a comparison of the peak luminosity with other black hole transients, such as \mbox{GX~339$-$4} and \mbox{MAXI~J1820+070}. Assuming the luminosity peak is sub-Eddington and at a similar level as in \mbox{MAXI~J1820+070}, \citet{Veledina2023} estimated the distance $D \approx 1.5$\,kpc. Assuming the spectral-fit parameters from Table~\ref{table:spectral}, we can estimate the current accretion rate in Eddington units as $L/L_{\rm Edd} = \eta \dot{M} c^2 / 1.26\times10^{31} M_{\rm bh}/M_{\odot}$, %which gives for the accretion efficiency $\eta \approx 0.15$ corresponding to the black hole spin $a = 0.88$, accretion rate $\dot{M} = 10^{17}$\,g\,s$^{-1}$ and black-hole mass $M_{\rm bh}=10.65 M_{\odot}$: $L/L_{\rm Edd} \approx 0.9\%$. 
which gives for the accretion efficiency $\eta \approx 0.14$ corresponding to the black hole spin $a = 0.87$, the accretion rate $\dot{M} = 10^{17}$\,g\,s$^{-1}$ and the black-hole mass $M_{\rm bh}=10 M_{\odot}$: $L/L_{\rm Edd} \approx 1\%$.

This is consistent with the soft-to-hard transition usually happening at $(0.5\%-4\%)L_{\rm Edd}$ \citep{Maccarone2003, Dunn2010}, while the luminosity at the hard-to-soft transition has a significantly larger scatter even for the same source \citep[e.g., in GX 339-4,][]{Dunn2008}. We tested the models with the lower distance estimate, which gave us a spin estimate $a \approx 0.95$, inclination $i \approx 48\degr$, and the accretion rate $\dot{M} \approx 2.8\times10^{16}$\,g\,s$^{-1}$, corresponding to $\approx 0.4\%$, just below the usual interval for the Eddington ratio at the transition. From these considerations, the larger distance with the accretion rate $\approx 1\%$ better matches the behavior of other black hole X-ray transients.

%\textbf{The outburst shape is also different than the ones from \mbox{GX~339$-$4} and \mbox{MAXI~J1820+070}, with the X-ray flux decreasing from the hard-state peak towards the soft state. Similar behavior was reported in several sources in the past, e.g., \mbox{GRS 1758$-$258} \citep{Smith2001}, \mbox{1E 1740.7$-$2942} \citep{delSanto2005}, and \mbox{4U1630$-$40} \citep{Tomsick2014}. The most likely explanation is a higher inclination of {\source}. The dependence of the outburst appearance on the inclination was analyzed by \citet{Munoz-Darias2013} for a {\rxte}-observed sample of 11 black-hole transients that reached a fully disk-dominated soft state. The differences could be explained by the relativistic effects on the accretion disk radiation.
%The angular distribution of radiation coming from the standard accretion disk and a hot flow is very different. In the soft state, the disk radiation is strongly beamed perpendicular to the disk, while in the hard state, the (optically thin) hot flow emission is more isotropic and beamed along the disk because of the Doppler boosting. Thus at a high inclination, the transition to the soft state can result in a decrease of the observed flux despite the increased accretion rate. The inferred high inclination also aligns with spectral-timing results obtained with \textit{Insight}-HXMT observations of {\source}, based on the measured QPO rms variability and phase lags \citep{yu2024timing}.} 
%The low-inclined sources have instead softer spectra and their accretion disks look cooler.

\section{Conclusions}
\label{sec:conclusions}

We report the detection of polarization changes correlated with the spectral transition in the low-mass black hole X-ray binary system \source.
% The X-ray polarization measurement of {\source} in a decaying soft state reveals how much the PD changes with the accretion state. 
Our new observations show that the PD substantially decreased from the previously measured 3\%--4\% to less than 1.2\% at the 99\% confidence level.
Such a drop in the PD favors the scenario in which the X-ray polarization signal is driven by the configuration of the accretion flow in the innermost region and indicates that the changes in spectral states are closely followed by the changes in polarization properties in a predicted way.
%Given the domination of the thermal disk component in the current state, the X-ray polarization signal is expected to change compared to the hard state when the X-rays are dominated by an up-scattered emission from a hot X-ray corona. 
%At the beginning of the hard-to-soft state transition, the PD was about $\approx 4\%$ \citep{Veledina2023}, it decreased to $\approx 3\%$ \citep{Ingram2024} towards the soft-intermediate state and dropped below $<1.2\%$ in the soft state. 
This substantial change of the PD was measured for the first time with {\ixpe} in the same source, and it indicates that the X-ray polarization is sensitive to the innermost accretion geometry. %and is significantly higher in the hard state, in which the X-ray corona up-scatters the accretion disk photons. 
The changes are in line with the early expectations that the radiation is produced in the optically thin medium in the hard state, switching to the optically thick medium in the soft state.
The upper limit of 1.2\% in the soft state, combined with the hard-state measurements, indicates that the inclination of the accretion disk is intermediate ($30\degr\lesssim i \lesssim 50\degr$).
%, similar to the range of values implied by optical spectroscopy and X-ray spectral modeling.
%The indicated PA is consistent with previous measurements in the hard state.
 
\section*{Acknowledgments}

IXPE is a joint US and Italian mission.  The US contribution is supported by the National Aeronautics and Space Administration (NASA) and led and managed by its Marshall Space Flight Center (MSFC), with industry partner Ball Aerospace (contract NNM15AA18C).  The Italian contribution is supported by the Italian Space Agency (Agenzia Spaziale Italiana, ASI) through contract ASI-OHBI-2022-13-I.0, agreements ASI-INAF-2022-19-HH.0 and ASI-INFN-2017.13-H0, and its Space Science Data Center (SSDC) with agreements ASI-INAF-2022-14-HH.0 and ASI-INFN 2021-43-HH.0, and by the Istituto Nazionale di Astrofisica (INAF) and the Istituto Nazionale di Fisica Nucleare (INFN) in Italy.  This research used data products provided by the IXPE Team (MSFC, SSDC, INAF, and INFN) and distributed with additional software tools by the High-Energy Astrophysics Science Archive Research Center (HEASARC), at NASA Goddard Space Flight Center (GSFC).
This research has used the MAXI data provided by RIKEN, JAXA, and the MAXI team.

J.S., M.D., J.Pod. and S.R.D. thank GACR project 21-06825X for the support and institutional support from RVO:67985815. 
A.V. thanks the Academy of Finland grant 355672 for support. Nordita is supported in part by NordForsk.
The work of F.M., R.T., G.M., L.M. and P.S. is partially supported by the PRIN grant 2022LWPEXW of the Italian Ministry of University and Research (MUR).
M.B. acknowledges the support from GAUK project No. 102323.
A.I. acknowledges support from the Royal Society. 
T.M.-D. acknowledges support by the Spanish \textit{Agencia estatal de investigaci\'on} via PID2021-124879NB-I00.
The French contribution is supported by the French Space Agency (Centre National d'Etude Spatiale, CNES) and by the High Energy National Programme (PNHE) of the Centre National de la Recherche Scientifique (CNRS).
Y.Z.\ acknowledges support from the Dutch Research Council (NWO) Rubicon Fellowship no.\ 019.231EN.021.
F.C. acknowledges the support of the INAF grant 1.05.23.05.06: “Spin and Geometry in accreting X-ray binaries: The first multi frequency spectro-polarimetric campaign".
Finally, we thank the anonymous Referee for the insightful comments which helped to improve the paper.

\appendix
\section{Details of spectral analysis}
\label{appendix-spectral}

Table~\ref{table:spectral} lists only free physical parameters of the spectral fit.
Besides the physical model components, we included a cross-calibration constant to account for calibration uncertainties between different instruments, an \texttt{edge} model to account for {\nicer} calibration feature around 2\,keV, and {\texttt{gain}} command to allow for the modification of response functions with the slope and offset parameters to be free among the {\ixpe} three detector units.
The final model in the \textsc{xspec} notation is: \texttt{const*edge*tbabs*(thcomp*kerrbb + mbknpo*kynxillver)}.

We limit the {\nicer} data to be above 1\,keV to avoid the need for further modeling of calibration uncertainties and more complex absorption features. The cross-normalization constants were fixed to 1 for {\nicer}, up to 5\% difference was allowed for {\nustar} and free for {\ixpe}. We obtained 0.95 (pegged) for {\nustar}-FPMA, 0.93 ($\pm 0.01$) for {\nustar}-FPMB, $0.92$ ($\pm 0.06$) for {\ixpe}-GPD1, $0.91$ ($\pm 0.06$) for {\ixpe}-GPD2, and $0.89$ ($\pm 0.06$) for {\ixpe}-GPD3. For the {\nicer} absorption edge, we got energy $E = 1.82 \pm 0.02$\,keV and the absorption depth $\tau = 0.03 \pm 0.01$. 
%The obtained gain parameters for {\ixpe} responses were: DET1: slope $s$ = 0.97 ($\pm 0.01$), and offset $o$ = 0.06 ($\pm 0.02$), DET2: $s$ = 0.96 ($\pm 0.01$), $o$ = 0.13 ($\pm 0.01$), and DET3: $s$ = 0.97 ($\pm 0.01$), $o$ = 0.11 ($\pm 0.01$). 
The obtained gain parameters for {\ixpe} responses were: DET1: slope $s$ = 0.97 ($\pm 0.01$), and offset $o$ = 0.05 ($\pm 0.01$), DET2: $s$ = 0.96 ($\pm 0.01$), $o$ = 0.11 ($\pm 0.01$), and DET3: $s$ = 0.97 ($\pm 0.01$), $o$ = 0.09 ($\pm 0.01$).

In the \texttt{kerrbb} model, we set the hardening factor to hd = 1.7 \citep{Shimura1995}. We employed limb darkening and self-irradiation, and fixed the distance to the value of 2.7\,kpc reported by \citet{Mata2024}, and fixed the normalization to unity. The electron temperature was set to 150\,keV in the {\texttt{thcomp}} model. The energies were extended to 0.01--100\,keV to properly calculate the Comptonization spectral component.
For the reflection model, we used the \texttt{xillver} tables \citep{Garcia2010} convolved with the relativistic model \texttt{KY} \citep{Dovciak2004}. We employed the lamp-post geometry model with the height set to 3 gravitational radii ($GM/c^{2}$), which has a similar irradiation profile like an extended corona with $r^{-3}$ \citep{Dovciak2014}. We used the standard low-density {\texttt{xillver}} tables with the density $10^{15}$\,cm$^{-3}$, but for the ionization calculations, we assumed the density to be $10^{18}$\,cm$^{-3}$. The gradient ionization over the disk radii dependent on illumination was assumed \citep{Svoboda2012}. 
The photon index of the reflection model was set to 3.4, which is the highest value of the pre-calculated \texttt{xillver} tables. The iron abundance was set to the solar one. The normalization was fixed to $1.37\times10^5$ (corresponding to $1/D^{2}_{\rm Mpc}$ as defined in the model).

Because the reflection models are calculated assuming the power law extending down to very low energies and thus overpredict the contribution of the reflection at $\lesssim 1$\,keV, we applied a correction for the low-energy part by multiplying the reflection model with a \texttt{mbknpo} model. This applies a break to the power-law shape of the reflection below a reference energy corresponding to some multiple of the thermal disk peak. Below this energy, the photon spectrum turns over which prevents the unphysical runaway.  From application of this model to several customized \texttt{xillver} reflection computations, an empirical scaling was adopted and here we force the low-energies to follow the slope $E^{-1.5}$, based on our empirical investigations of the shape of reflection models illuminated by the {\texttt{nthcomp}} model.
In the \textsc{xspec} notation, we defined this function as: \texttt{mdef mbknpo} $({\rm max}(E,B)-B)/{\rm abs}(E-B+\epsilon)+(1-({\rm max}(E,B)-B)/{\rm abs}(E-B+\epsilon))*(E/B)^I$: \texttt{mul}, where $B$ is the break energy, $I$ is the index used for the correction for energies $E<B$, $\epsilon$ (set to $10^{-7}$) is just a small number to avoid divergences in the numerical calculation of the model at $E = B$.
The index below the break energy was set to $I = \Gamma - 1.5 = 1.9$, where $\Gamma$ is the photon index of illuminating power law in the reflection model. The value of the break energy was fitted and found to be $B = 2.7 \pm 0.2$~keV.

\bibliography{refs}
\bibliographystyle{aasjournal}

\end{document}